\documentclass[journal,comsoc]{IEEEtran}

\usepackage[T1]{fontenc} 
\usepackage{cite}

\usepackage[hidelinks]{hyperref}

\ifCLASSINFOpdf
    \usepackage[pdftex]{graphicx}
    \usepackage{subfig}
    \DeclareGraphicsExtensions{.pdf,.jpeg,.png}
\else
\fi
\usepackage{amsmath}

\usepackage[cmintegrals]{newtxmath}
\usepackage{array}
\usepackage{longtable}
\usepackage{rotating}

\usepackage{stfloats}
\usepackage{url}

\usepackage{multirow}
\usepackage{array}
\usepackage{booktabs}

\usepackage{multirow}
\usepackage{enumerate}
\usepackage[linesnumbered,ruled,algo2e] {algorithm2e}
\usepackage{algorithm}
\usepackage{algorithmic}
\usepackage{verbatim}

\usepackage{xargs}
\usepackage[pdftex,dvipsnames]{xcolor}
\usepackage[colorinlistoftodos,prependcaption,textsize=tiny]{todonotes}
\usepackage[section]{placeins}
\newcommandx{\unsure}[2][1=]{\todo[disable,linecolor=red,backgroundcolor=red!25,bordercolor=red,#1]{#2}}
\newcommandx{\change}[2][1=]{\todo[disable,linecolor=blue,backgroundcolor=blue!25,bordercolor=blue,#1]{#2}}
\newcommandx{\info}[2][1=]{\todo[disable,linecolor=OliveGreen,backgroundcolor=OliveGreen!25,bordercolor=OliveGreen,#1]{#2}}
\newcommandx{\improvement}[2][1=]{\todo[disable,linecolor=Plum,backgroundcolor=Plum!25,bordercolor=Plum,#1]{#2}}

\begin{document}

\title{A Solution of Ultra Wideband Based High-resolution and Lossless Audio Transmission
\thanks{Fengyun Zhang is with the College of Artificial Intelligence, Southwest University, Chongqing, China (email: zhangfy2019@mail.sustech.edu.cn).}
}

\author{Fengyun~Zhang,~\IEEEmembership{Member,~IEEE / AES}}


\markboth{IEEE Internet of Things Journal, ~Vol.~xx, No.~x, Aug~2025}%
{}


\maketitle

\begin{abstract}

This paper provides an overview of the current challenges in wireless audio transmission and highlights the limitations of existing technologies regarding data bandwidth, data compression, latency, and inter-device compatibility. To address these shortcomings, it proposes a high-resolution, lossless audio transmission scheme utilizing ultra wideband (UWB) technology. UWB emerges as a promising solution by offering the necessary bandwidth to enable exceptional sound quality with ultra-low latency, making it ideal for real-time audio applications and addressing synchronization concerns in audio-visual use cases. Additionally, UWB's unique capabilities extend beyond high-resolution audio, allowing for precise location tracking in augmented and virtual reality applications.  
\end{abstract}

\begin{IEEEkeywords}
Ultra wideband (UWB), wireless audio transmission, high-resolution and lossless
\end{IEEEkeywords}

\IEEEpeerreviewmaketitle

\section{Introduction}
\label{sec:introduction}
\IEEEPARstart{I}{n} the realm of consumer electronics, achieving high-quality audio transmission has always been a fundamental goal. Various technologies have been developed to meet this objective. The UWB Alliance joined the Audio Engineering Society (AES) Task Group in November 2023, focusing on developing a new standard for UWB high-resolution, low-latency audio interfaces \cite{keyuwba}. This initiative is chaired by the UWB Alliance and aims to enhance high-quality audio applications. The inaugural meeting of the Audio Engineering Society to formulate this object represents a significant advancement in achieving superior audio performance \cite{keyaes}. Table~\ref{tab:table-1} presents a comparison of the advantages and limitations of different wireless protocols. Bluetooth \cite{8721261} is widely used and has good device compatibility and low-power, making it a popular choice. However, Bluetooth has limitations in terms of audio quality. Wi-Fi \cite{4147519} offers higher bandwidth and enables high-quality audio transmission. Compared to Bluetooth, WiFi is capable of transmitting uncompressed audio data, resulting in superior audio quality. There are also other wireless protocols to consider, such as DLNA (Digital Living Network Alliance) \cite{5456206}, AirPlay \cite{richter2013airplay}, Chromecast \cite{keychromecast}, and UWB \cite{9179124}. 
\begin{table}[htb]
	\caption{Advantages and limitations of different wireless protocols.}
	\centering
	\begin{tabular}{  m{1.2 cm}  m{3.0 cm} m{3.0 cm}}
		\hline \specialrule{0em}{2pt}{2pt} 
		\textbf{Approach} & \textbf{Advantages} & \textbf{Limitations} \\ \specialrule{0em}{2pt}{2pt} \hline \hline \specialrule{0em}{2pt}{2pt} 
		Bluetooth & Widely used, good device compatibility, and low-power. &Limited audio quality, when high bit rates or high sampling rates audio. \\ \hline \specialrule{0em}{2pt}{2pt}
		Wi-Fi & Higher bandwidth, high-quality audio transmission & Not well-suited for mobile audio devices. \\ \hline \specialrule{0em}{2pt}{2pt} 
		DLNA & Multi-device connectivity and high-quality audio transmission. & Only DLNA-compatible devices. \\ \hline \specialrule{0em}{2pt}{2pt} 
		AirPlay & High-quality audio transmission, and good compatibility with Apple devices. & Limited to the Apple ecosystem. \\ \hline \specialrule{0em}{2pt}{2pt} 
  	Chromecast & Lossless and high-quality audio and video transmission. & Limited to the Google ecosystem. \\ \hline \specialrule{0em}{2pt}{2pt} 
		UWB & No bandwidth limitations, and lossless audio and video transmission. & A currently limited number of supported devices. \\ \specialrule{0em}{2pt}{2pt} \hline
	\end{tabular}
	\label{tab:table-1}
\end{table}

The following key points are essential for accomplishing high-resolution and lossless audio transmission, including high bandwidth, Lossless Audio Files, Codec Technology, Signal Quality, Device Compatibility, Network Stability, and Proper Configuration. To attain high-resolution lossless audio transmission, the selection of wireless transmission technology becomes especially crucial, particularly for wireless mobile audio devices sensitive to delay and power consumption. This paper aims to provide an solution of high-resolution and lossless audio transmission based on UWB, highlighting its strengths and limitations. The main contributions of this paper can be summarized as follows: 
\begin{itemize}
    \item We summarize existing wireless protocols, audio codecs, and Bluetooth codecs, providing a comprehensive overview of the current state of the field.
    \item We innovatively introduce UWB technology as a solution for high-resolution and lossless audio transmission. 
    \item We emphasize the advantages of UWB as audio transmission technology, and also point out the trend of UWB development in the future.  
\end{itemize}


\section{Solution Analysis}
\label{sec:solution_analysis}
Audio encoding falls into two categories: lossy and lossless, with the former causing some audio quality reduction. In theory, no digital audio format can perfectly replicate an analogue signal. PCM encoding initiates the conversion of analogue to digital with the highest fidelity, often referred to as "lossless coding." Other codecs are variations of PCM, designed to compress it, known as compression codecs. These established standards and formats, which are based on precise audio encoding algorithms and data structures, play vital roles in various domains such as audio systems, high-fidelity audio players, and audio transmission devices. They ensure the preservation of high-fidelity audio quality and enable efficient transmission and storage, which are critical for high-resolution lossless audio. The results for CD resolution and Hi-res audio are presented in Table~\ref{tab:1-2}.
\begin{table*}[htbp]
	\caption{Different levels of audio quality}\label{tab:1-2}
	\centering
	\begin{tabular}{ccccccc}
	\toprule
    Format & Bit-rate & Bit-depth & Sampling-rate & No of Channels &  Audio Bandwidth & Max SNR\\
	\midrule
    CD resolution & 1.411~mbps & 16-bit & 44.1~kHz & 2 & <~22~kHz & 96dB \\
    HD Audio & 2.304~mbps &  24-bit & 48~kHz  & 2 & <~24~kHz & 144dB \\
    Hi Resolution  & 4.608~mbps & 24-bit & 96~kHz & 2 & <~48~kHz & 144dB  \\
	Hi-Res Audio & 2.304~mbps &  24-bit  & 48~kHz & 2 & <~24~kHz & 144dB \\
    Tidal MQA & 4.608~mbps & 24-bit & 96~kHz & 2 & <~48~kHz & 144dB \\
    Apple Music Hi-Res & 9.216~mbps &  24-bit & 192~kHz & 2 & <~96~kHz & 144dB \\
    Amazon Music Ultra HD & 9.216~mbps &  24-bit  & 192~kHz & 2 & <~96~kHz & 144dB \\
	\bottomrule
	\end{tabular}
\end{table*} 

\subsection{Codecs of Lossless Audio Transmission}
\label{subsec:codec_lat}
High-resolution lossless audio transmission is a critical aspect of audio technology that seeks to ensure the transmission and storage of audio data without any loss in quality. Several key standards and formats in high-resolution lossless audio transmission have been established to facilitate the preservation of audio quality \cite{randhawa2022ieee, van2024rfc, van2022free}. PCM encoding, known as "lossless codec," offers the highest fidelity in analog-to-digital conversion. Other audio codecs created with PCM are as follows:
\begin{itemize}
	\item \textbf{FLAC (Free Lossless Audio Codec)}: FLAC is widely used for lossless audio transmission and storage. 
	\item \textbf{ALAC (Apple Lossless Audio Codec)}: ALAC is a lossless audio encoding format mainly employed in Apple devices. 
    \item \textbf{WAVpack}: WAVpack is a lossless audio encoding format that supports high-fidelity audio transmission. 
    \item \textbf{PCM (Pulse Code Modulation)}: PCM is a basic lossless audio encoding method typically used for digital storage of audio, such as in CDs.
    \item \textbf{MQA (Master Quality Authenticated)}: MQA compresses high-resolution audio while preserving audio quality. 
   \item \textbf{WAV (Waveform Audio File Format)}: WAV is a lossless audio file format commonly used to store uncompressed audio data. 
   \item \textbf{AIFF (Audio Interchange File Format)}: AIFF, similar to WAV, is also a lossless audio file format commonly used in Mac OS.  
   \item \textbf{APE (Audio File Format of Monkey’s Audio)}: Monkey’s Audio is a fast and easy way to compress digital music, absolutely no quality loss.  
\end{itemize}

\subsection{Audio Codecs of Bluetooth}
\label{subsec:codec_ble}
Wireless devices provide convenience in our daily lives, with Bluetooth being a prominent audio transmission standard. Achieving high-quality audio in this domain remains a challenge. Formats like FLAC or ALAC, known for lossless audio, are not supported by Bluetooth. In the realm of wireless audio, lossless audio transmission necessitates meeting three prerequisites: a high sampling rate, ample bit depth, and a substantial bit rate. It is worth noting that CD quality serves as the benchmark, supported by the conventional 16-bit audio data bit depth and a 44.1 kHz sampling rate. The minimum bit-rate required to transmit CD-quality audio data is 1411 Kbps ($44.1~kHz \times 16~bits \times 2 = 1411.2~kbps$) with 2 channels, not including the additional bandwidth needed for error correction. We will delve into various Bluetooth codecs and explore their distinctions in Table \ref{tab:1-3}, including sampling rate, bit-depth, and bit-rate. 
\begin{table}[htbp]
	\caption{Comparison of different codecs of Bluetooth}\label{tab:1-3}
	\centering
	\begin{tabular}{cccc}
	\toprule
    Codec & Bit-rate & Sampling rate & Bit-depth \\
	\midrule
    \textbf{LC3plus} & 128-672~kbps & 48 / 96~kHz & 24-bit  \\
    \textbf{L2HC (2.0/3.0/4.0)} & Up to 23K~kbps & 48~kHz & 24-bit  \\
    \textbf{SCL6 (MQair)} & 200-20K~kbps & 96 / 384~kHz & 24-bit  \\
    \textbf{LDAC}  &330-990~kbps & 192~kHz & 24-bit  \\
    \textbf{LHDC (platinum)} & 400-900~kbps &  96~kHz & 24-bit  \\
	aptX Adaptive  & 279-420~kbps & 48 / 96~kHz & 24-bit \\
	aptX HD  & 576~kbps  & 48~kHz & 24-bit  \\
	aptX & 325 / 384~kbps & 44.1 / 48~kHz  & 16-bit   \\
	AAC  & 320 / 512~kbps  & 44.1 / 48~kHz & 16-bit \\
	SBC & 328~kbps & 44.1~kHz &  16-bit  \\
    LC3 & 160-345~kbps & 8-48~kHz  & 16-bit  \\
	\bottomrule
	\end{tabular}
\end{table}

To obtain the certification of "Hi-Res Audio Wireless", a codec must support a frequency range of up to 40 kHz and support 24-bit/96~kHz. Currently, only five codecs, LDAC, LHDC, LC3plus, L2HC, and SCL6 meet these stringent standards. The SCL6 exhibits significant bitrate flexibility, ranging from a maximum of 20~mbps to a minimum of 200~kbps. However, it is important to note that current Bluetooth connections, including the latest Bluetooth 5.3, are incapable of supporting bit rates approaching 2.0~mbps. Even well-established LDAC and LHDC codecs, which claim bit rates of up to 990~kbps and 900~kbps, encounter challenges in sustaining these rates during continuous operation, frequently experiencing drops to lower rates. L2HC based on the nearlink technology is currently only used for Huawei's audio products.

UWB technology is poised to revolutionize audio transmission, offering high-quality, low-latency, secure, and interference-resistant data transmission services. However, the current lack of standardized audio specifications cannot be overlooked. An Overview of UWB Standards and Organizations could be found in reference\cite{9810941}, and a summary of the main UWB hardware solutions available today is presented in Table~\ref{tab:1-4}. For the vast majority of UWB technology solution providers, the initial design of UWB chips did not specifically target audio transmission, with their application scenarios primarily focused on ranging and IoT domains. The UWB chip introduced by SPARK can be utilized for audio transmission, but it has not received FiRa (FiRa Consortium) certification\cite{fira}. CEVA, a globally recognized leader in IP core licensing, specializes in UWB technology for audio transmission. However, it does not engage in the independent manufacturing of chips \cite{ceva}. Given the widespread application in smartphones, the UWB chip launched by Qualcomm in 2024 is most likely to become the dominant player in the next generation of lossless audio transmission technology\cite{qualcomm}.
\begin{table*}[htbp]
	\caption{Comparison of different Ultra Wideband solutions}\label{tab:1-4}
	\centering
	\begin{tabular}{cccccc}
	\toprule
    Company & Device Name & Consortium  & Features & Standard & Band \\
	\midrule
    Qorvo\cite{qorvo} & DW3220 / DW3720 / DW35725 / QM33120W & FiRa &  Range / Radar / IoT / Smartphone & HRP &  3.5-8.5 GHZ\\
    NXP\cite{nxp} & SR150 / SR040 / SR100T &  FiRa &  Range / Radar / IoT / Smartphone & HRP & 6.0-9.0 GHz\\
    Qualcomm & WCN788x & FiRa  & Range / Audio / IoT / Smartphone& HRP & 6.0 GHz\\
    Apple\cite{apple} & Apple U1 / U2 & FiRa & Rang / IoT / Smartphone & HRP &  6.0–8.5 GHz\\    
    CEVA & IP Core &  FiRa   & Range / Radar / Audio / IoT & HRP & 3.1–10.6 GHz \\
    SPARK & SR1010 / SR1020 & FiRa  & Range / Audio / IoT & N/A & 3.1–9.25 GHz \\
    NewRadioTech\cite{newradiotech} & NRT82885 / NRT81750 &  FiRa  & Range / Radar / IoT & HRP & 6.0-9.0 GHz\\
    CXSEMI\cite{cxsemi} & CX310 / CX500 &  FiRa  & Range / Radar / IoT & HRP & 6.0-9.0 GHz\\
    GiantSemi\cite{giantsemi} & GT1500 / GT1000 &  FiRa   & Range / Radar / IoT / Smartphone & HRP & 4.0-9.0 GHz\\
	\bottomrule
	\end{tabular}
\end{table*} 

\subsection{Advantages and Challenges of UWB for Lossless Audio Transmission}
\label{subsec:pcc_uwb}

\textbf{Advantages of UWB in audio transmission: }
\begin{itemize}
    \item \textbf{High bandwidth}: UWB provides an extensive bandwidth capable of transmitting large volumes of data, including high-quality audio. 
    \item \textbf{Low Latency}: UWB typically offers minimal transmission latency, making it suitable for real-time audio applications, such as wireless audio headphones.
   \item \textbf{Interference resilience}: UWB excels at handling multipath transmission, reducing signal reflections and interference, thereby improving signal quality.
    \item \textbf{Low power consumption}: UWB efficiently operates by transmitting high bandwidth data in short bursts, allowing for lower average power consumption.
\end{itemize}

\textbf{Challenges of UWB in audio transmission: }
\begin{itemize}
    \item \textbf{Bandwidth management}: Optimization for bandwidth management, encompassing enhanced spectrum allocation and interference mitigation strategies. 
    \item \textbf{latency and synchronization}: UWB technology necessitates a low-latency, highly precise clock synchronization mechanism to ensure synchronous audio playback across diverse devices.
    \item \textbf{Advanced antenna design}: The issue arises from the reflection or absorption of radio waves generated by existing UWB antennas by the human body, thereby causing signal interference.
    \item \textbf{Multipath interference suppression}: Sophisticated signal processing algorithms must be incorporated to mitigate multipath interference and uphold sound quality. 
    \item \textbf{Multi-device interoperability}: A unified UWB technology standard is essential to streamline device connectivity and collaboration, enhancing device interoperability. 
\end{itemize} 

\textbf{The primary factors of UWB will be the future in consumer electronic products. }
\begin{itemize}
    \item \textbf{UWB audio standard}: Wireless communication standards play a critical role in the advancement of UWB technology. Organizations such as IEEE, AES, and the UWB Alliance are actively accelerating the development of UWB as a standard for audio transmission. 
    \item \textbf{Ultra-low latency}: UWB technology enables ultra-low-latency audio transmission, while precise clock synchronization maintains audio data synchronization.
    \item \textbf{Ultra-low Power}: UWB technology achieves ultra-low power consumption communication, with energy efficiency as low as 1.5 nJ/bit.
    \item \textbf{Security and privacy}: Emphasizing audio security and privacy, proprietary encryption and authentication are used to protect audio data. 
\end{itemize} 

\section{UWB-based Audio Transmission Solution}
\label{sec:solution_UWB_Audio}

\subsection{Framework and Architecture of UWB-based Audio Transmission}
\label{subsec:Framework_UWB_Audio}
\begin{figure}[htbp]
	\centering
	\includegraphics[width=0.4\textwidth]{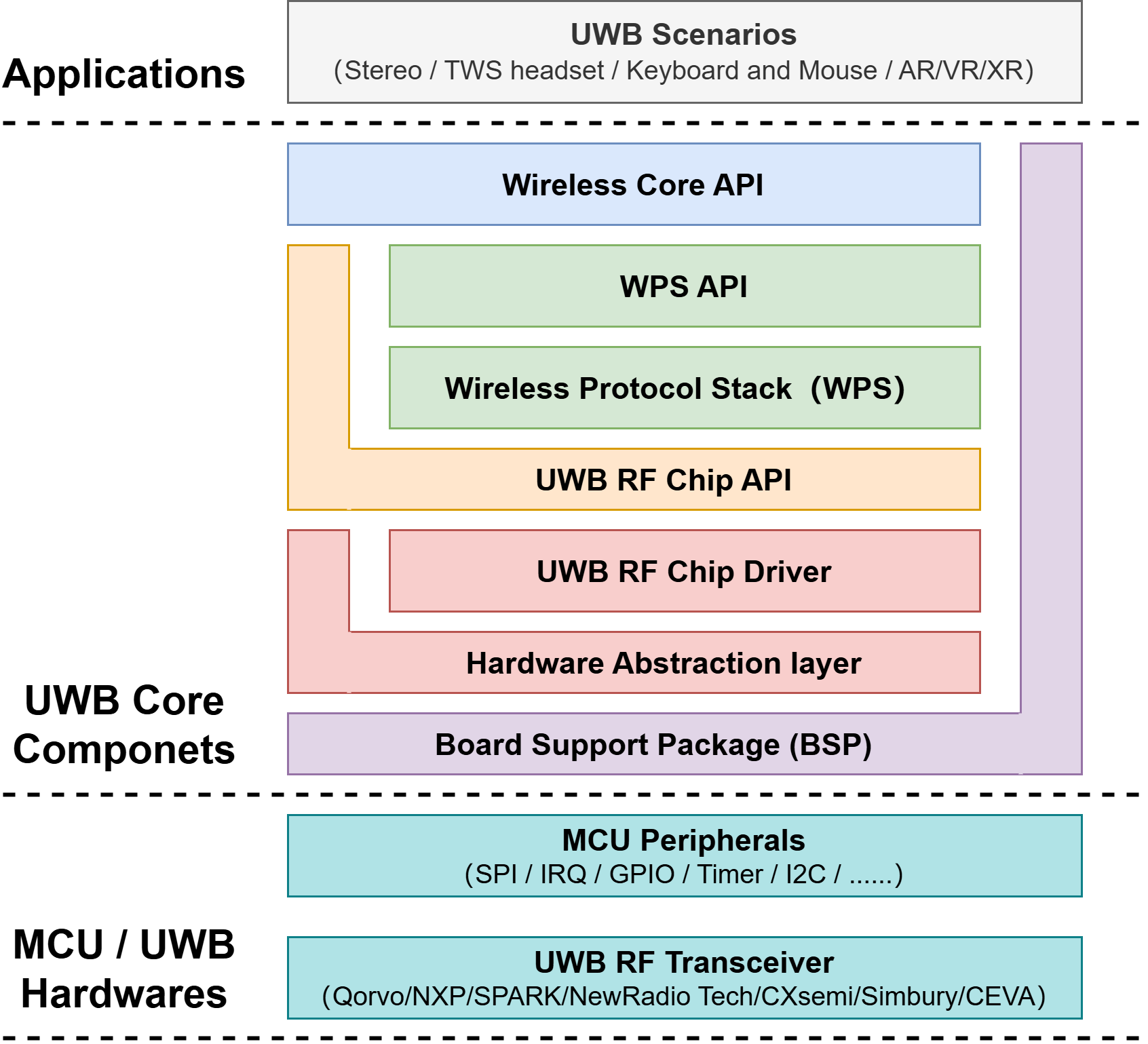}
	\caption{Framework of UWB-based audio transmission.}
	\label{fig:fig1}
\end{figure}

Through a comparative analysis of lossless audio transmission technologies and codecs, we outline a generalized solution for UWB-based lossless audio transmission, as shown in Figure~\ref{fig:fig1}. The UWB-based audio transmission framework primarily consists of three components: application scenarios, core components, and MCU/UWB hardware. The core component plays a crucial role in UWB technology for efficient audio transmission. Within this core component, we design and implement a proprietary wireless protocol stack (WPS). By leveraging the WPS, we achieve high-resolution, lossless audio transmission using UWB technology, such as HiFi stereo and TWS headset.

\begin{figure}[htbp]
	\centering
	\includegraphics[width=0.4\textwidth]{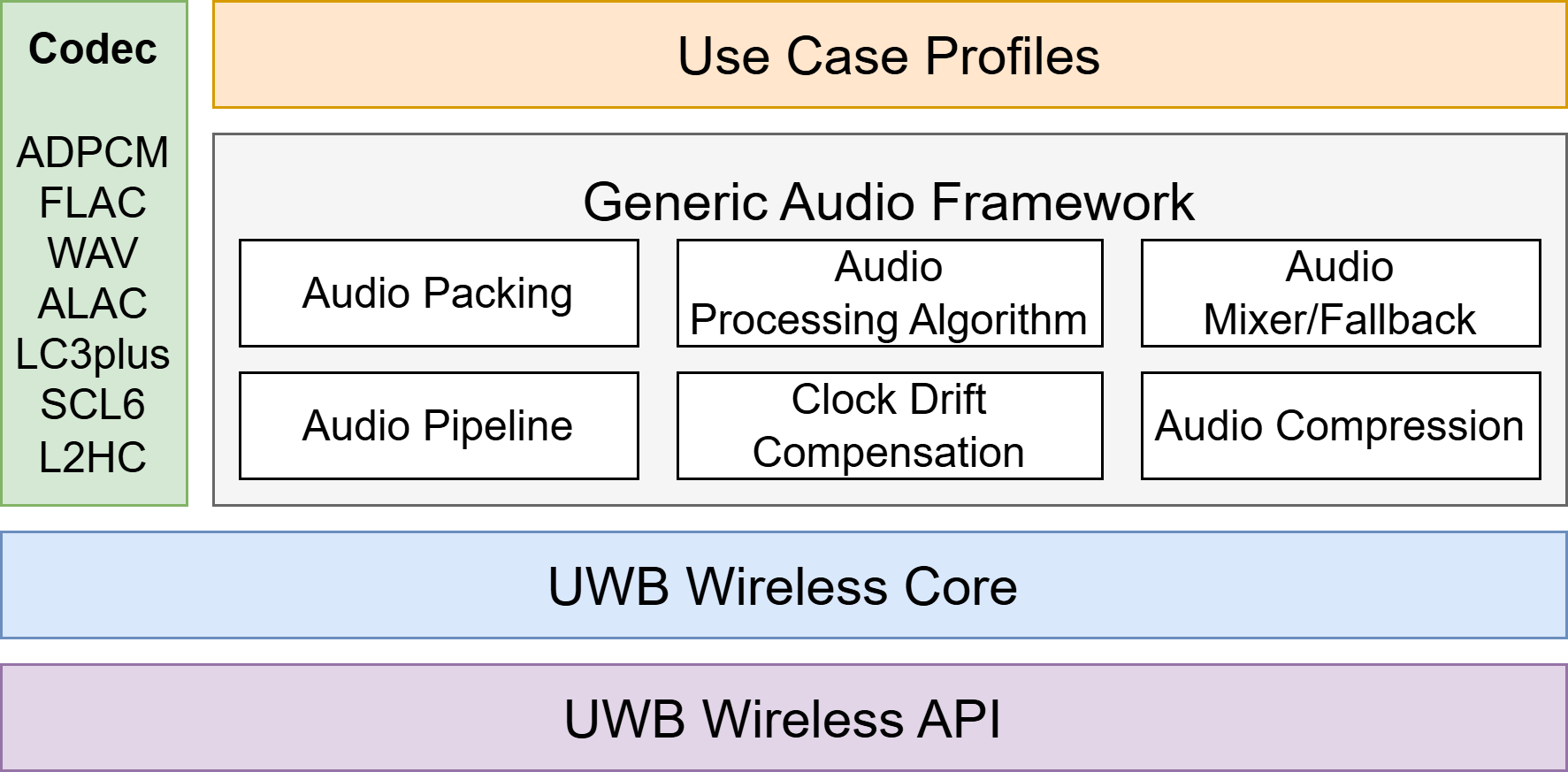}
	\caption{Architecture of UWB-based audio core.}
	\label{fig:fig2}
\end{figure}

The UWB-based audio core is the software utilized by all audio applications, and its architecture is illustrated in Figure~\ref{fig:fig2}. The primary function of the audio core is to manage audio samples, encompassing a generic audio framework, typical codecs, and the UWB wireless core and API. We aim for the UWB-based audio core to not only support ADPCM audio transmission but also to be compatible with other lossless audio codecs. The general audio framework consists of six main components: audio packing, audio processing algorithms, audio mixer/fallback, audio pipeline, clock drift compensation, and audio compression.

\begin{figure}[htbp]
	\centering
	\includegraphics[width=0.5\textwidth]{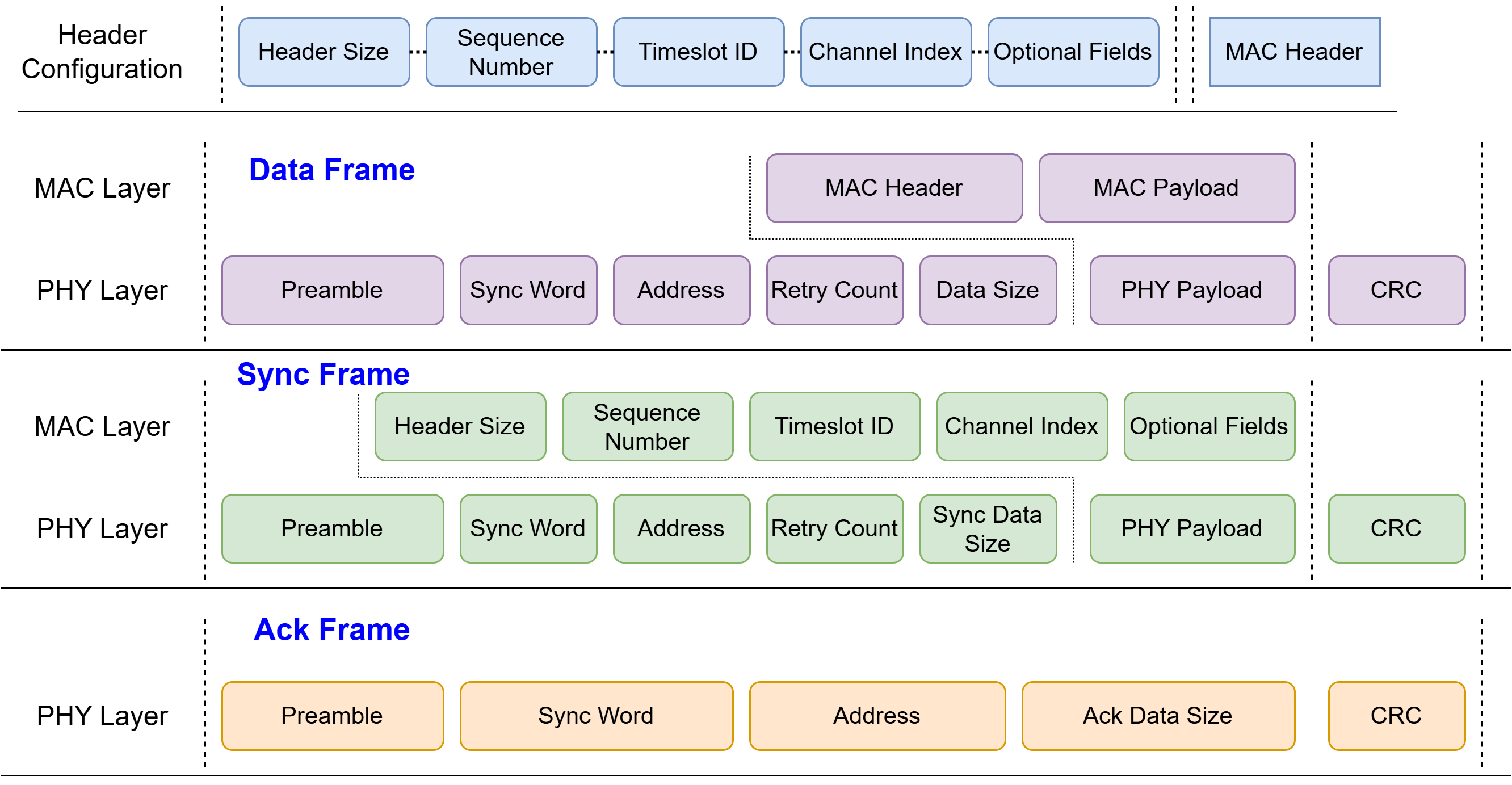}
	\caption{Frame Structure of wireless core.}
	\label{fig:fig3}
\end{figure}

Based on the characteristics of audio data stream transmission, we have redesigned the frame structure in the Wireless core to ensure continuous and efficient audio data transmission over UWB, as demonstrated in Figure~\ref{fig:fig3}. The Wireless core supports three types of frames: data frames, synchronization frames (Sync frame), and acknowledgment frames (Ack frame). It is important to note that certain bytes in the physical layer payload are reserved for use in the MAC layer. Data frames carry an application payload and can be transmitted as main frames or in an auto-reply mode. Synchronization frames are utilized when auto-sync mode is enabled on a connection. If the connection queue is empty, a frame containing only a header is sent to maintain network synchronization. Acknowledgment frames, which do not contain a payload, are exclusively sent in an auto-reply. This structure ensures reliable and efficient communication within the network.

\begin{figure}[htbp]
	\centering
	\includegraphics[width=0.5\textwidth]{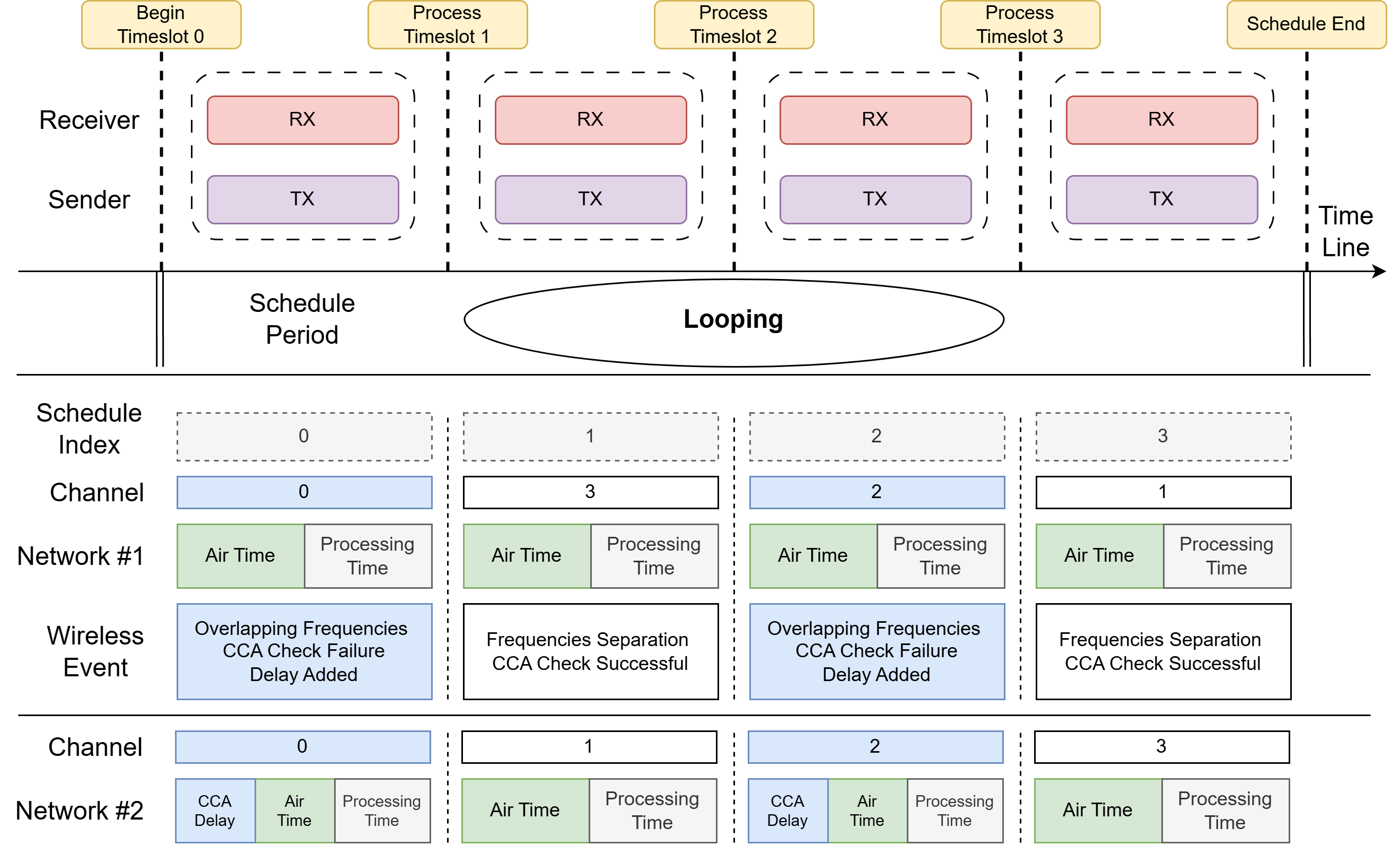}
	\caption{Scheduling method based on wireless protocol stack.}
	\label{fig:fig4}
\end{figure}

In the design of bandwidth scheduling within the wireless protocol stack, it is essential to consider the bandwidth requirements of each device. The peak rate of a connection can be determined based on the number of timeslots allocated within a specific time period, as illustrated in Figure~\ref{fig:fig4}. Synchronization timeslots, which consist of normal frames containing headers and payloads or empty frames containing only headers, play a critical role in maintaining synchronization. The maximum transmission period is set at 10 ms; exceeding this interval can lead to synchronization loss between the sender and the receiver. If the sender has no data to transmit during the synchronization timeslot, the auto-sync function can be utilized. This feature allows the wireless kernel to automatically manage the transmission of synchronized frames. During the first timeslot, the transmitter in Network 1 performs a Clear Channel Assessment (CCA) and determines that the air link is unobstructed, enabling it to start transmitting on channel 0. Concurrently, the transmitter in Network 2 also performs a CCA but fails due to the recent use of channel 0 by Network 1's transmitter. Consequently, the Network 2 transmitter waits for a predetermined delay before retrying. In the second time slot, both transmitters successfully transmit simultaneously as their CCA attempts both succeed.

\subsection{Evaluation of UWB-based Audio Transmission}
\label{subsec:evalutaion_UWB_Audio}
\begin{figure}[htbp]
	\centering
	\includegraphics[width=0.5\textwidth]{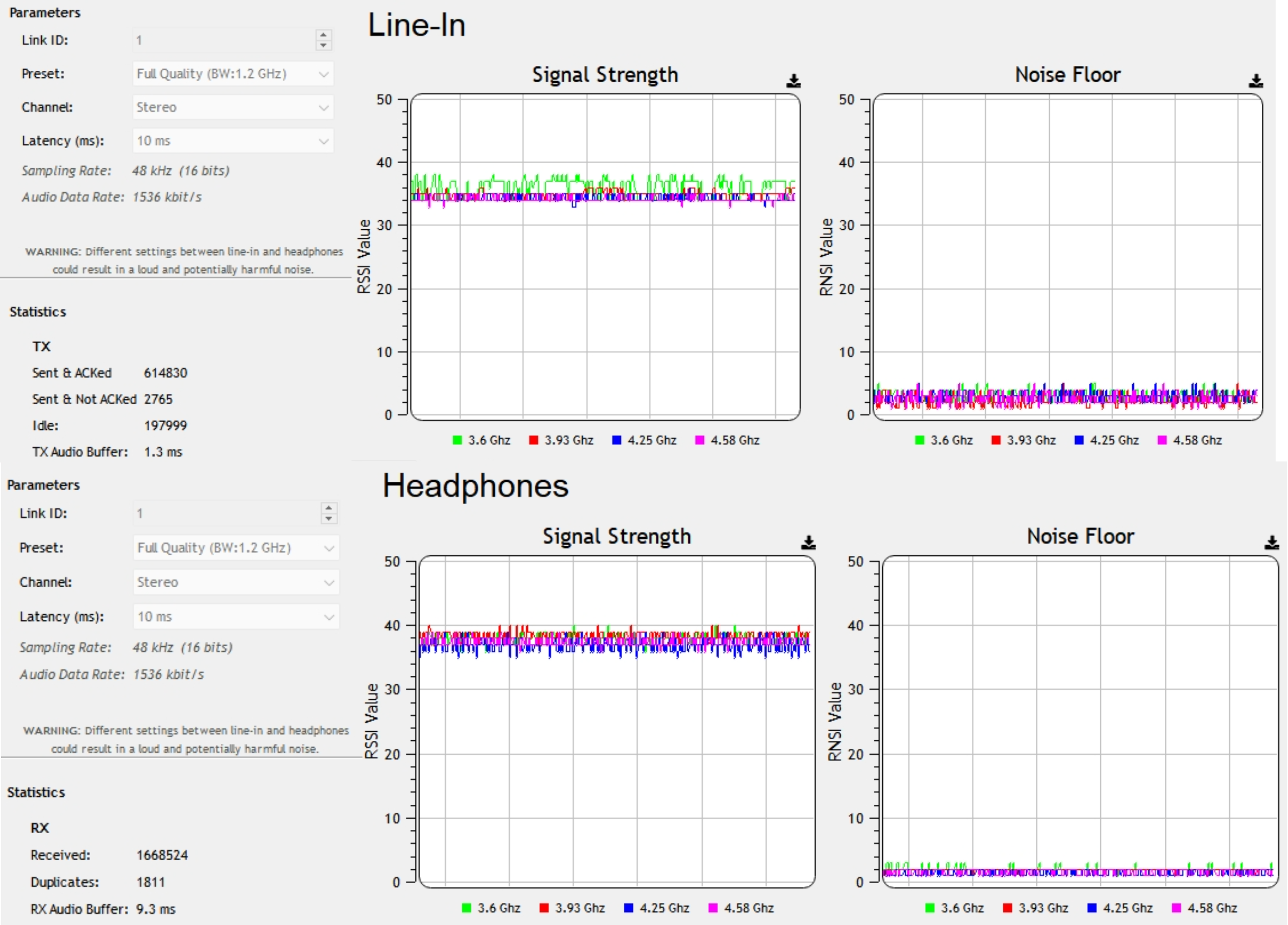}
	\caption{Evaluation results of UWB-based audio transmission.}
	\label{fig:fig5}
\end{figure}

According to the solution designed in Section~\ref{subsec:Framework_UWB_Audio}, we designed the lossless audio transmission experiment based on UWB, and the experimental results are shown in Figure~\ref{fig:fig5}. One evaluation board is connected to the lossless audio source via the Line-in port, while the other evaluation board is connected to the speaker through the headphone port. The operating mode of the development kit is configured for stereo, with a sampling rate of 48 kHz, a bit-rate of 1536 kbps, a bit-depth of 16, and a latency of 10 ms set through the host computer software. The experimental results demonstrate that our proposed lossless audio transmission scheme based on UWB can reliably transmit audio with a high signal-to-noise ratio (SNR). 

We conducted tests on actual communication latency and link utilization under different preset latencies at two bandwidths of 1.2GHz and 1.6GHz, as shown in Table~\ref{tab:1-5}. The statistical results indicate that our UWB solution, in full quality mode, generally exhibits communication latency lower than the preset delays, with link utilization around 76\%, and a communication success rate exceeding 99.9\%.
\begin{table*}[htbp]
	\caption{Comparison of different audio parameters}\label{tab:1-5}
	\centering
	\begin{tabular}{ccccccc}
	\toprule
    Parameter & Test Group 1 & Test Group 2 & Test Group 3 & Test Group 4 & Test Group 5 & Test Group 6 \\
	\midrule
    Bandwidth & 1.2 GHz & 1.2 GHz & 1.2 GHz  & 1.6 GHz & 1.6 GHz & 1.6 GHz\\
    Channel & Stereo & Stereo & Stereo  & Stereo & Stereo & Stereo\\
    Sampling rate & 48 KHz & 48 KHz & 48 KHz  & 48 KHz & 48 KHz & 48 KHz\\
    Bit-depth & 16 bits & 16 bits & 16 bits  & 16 bits & 16 bits & 16 bits\\
    Bit-rate & 1536 kbps & 1536 kbps & 1536 kbps  & 1536 kbps & 1536 kbps & 1536 kbps\\
    Preset latency  & 5 ms & 10 ms & 20 ms  & 5 ms & 10 ms & 20 ms \\
    \textbf{Real latency}  & 4.8 ms & 9.3 ms & 19.7 ms  & 4.7 ms & 9.5 ms & 19.3 ms \\
    \textbf{Link utilization} & 76\% &  77\% & 75\% & 76\% &  76\% & 75\% \\
    \textbf{Success rate} & 99.97\% &  99.99\% & 100\%  & 99.98\% &  99.99\% & 100\% \\
	\bottomrule
	\end{tabular}
\end{table*}

\section{Conclusion}
\label{sec:conclusion_and_discussion}
This paper addresses the key challenges in wireless audio transmission and identifies UWB technology as a promising solution for lossless audio. Moreover, An audio transmission solution based on UWB is proposed and its performance is evaluated. UWB offers the required bandwidth for lossless audio transmission and has remarkably low latency, addressing synchronization concerns in audio-visual scenarios. With its multi-device connectivity and enhanced functionality, UWB is poised to transform the consumer electronics landscape, enabling immersive and seamless audio experiences. 


%

\bibliographystyle{IEEEtran}
\bibliography{myrefs}

%
%

\ifCLASSOPTIONcaptionsoff
  \newpage
\fi

\end{document}